\documentstyle[12pt,psfig]{article}

\input epsf.tex
\newtheorem{thm}{Theorem}

\newtheorem{lem}{Lemma}
\newtheorem{rem}{Remark}

\newtheorem{exm}{Example}

\newcommand{\thmref}[1]{Theorem~{\rm \ref{#1}}}
\newcommand{\lemref}[1]{Lemma~{\rm \ref{#1}}}

\newcommand{\exmref}[1]{Example~{\rm \ref{#1}}}

\newcounter{neweqn}

\newcommand{\beq}[1]{\begin{equation} \refstepcounter{neweqn} \label{#1}}
\newcommand{\eeq}{\end{equation}}
\newcommand{\bed}{\begin{displaymath}}
\newcommand{\eed}{\end{displaymath}}
\newcommand{\bedd}{\bed\begin{array}{l}}
\newcommand{\eedd}{\end{array}\eed}
\newcommand{\nd}{\noindent}

\newcommand{\disp}{\displaystyle}
\newcommand{\A}{{\cal A}}

\newcommand{\rr}{{\hbox{{\rm I}{\kern -0.22em}{\rm R}}}}
\newcommand{\bdd}{\hspace*{-0.08in}{\bf.}\hspace*{0.05in}}

\newcommand{\Ii}{I_{\{\tau^b_1<\tau_M\}}}

\newcommand{\In}{I_{\{\tau^b_n<\tau_M\}}}

\newcommand{\IIn}{I_{\{\tau^{b*}_n<\tau_M\}}}

\def\({\left(}
\def\){\right)}

\begin{document}

\title{An Optimal Pairs-Trading Rule\thanks{This research is supported in part by the Research Grants Council of Hong Kong               
No. CityU 103310 and in part by the Simons Foundation (235179).}}

\author{Qingshuo Song\thanks{Department of Mathematics,
City University of Hong Kong, 83 Tat Chee Ave, Kowloon, Hong Kong,
song.qingshuo@cityu.edu.hk}
\and
Qing Zhang\thanks{Department of Mathematics, University of Georgia,
Athens, GA 30602, qingz@math.uga.edu}
}


\maketitle

\begin{abstract}
This paper is concerned with a pairs trading rule.
The idea is to monitor two historically correlated securities. 
When divergence is underway, i.e., one stock moves up while the other 
moves down, a pairs trade is entered which consists of a pair to 
short the outperforming stock and to long the underperforming one.
Such a strategy bets the ``spread'' between the two would eventually 
converge.
In this paper, a  difference of the pair is governed by a mean-reverting
model. The objective is to trade the pair so as to
maximize an overall return. A fixed commission cost is charged with each
transaction. In addition, a stop-loss limit is imposed as a state 
constraint.
The associated HJB equations (quasi-variational
inequalities) are used to characterize the value functions. It is
shown that the solution to the optimal stopping problem
can be obtained by solving a number of quasi-algebraic equations.
We provide a set of sufficient conditions in terms of 
a verification theorem. 
Numerical examples are reported to demonstrate the results.


\end{abstract}

\newpage

\baselineskip25pt
\renewcommand{\arraystretch}{1.8}

\section{Introduction}

This paper is concerned with pairs trading.
The idea is to identify and monitor a pair of historically correlated stocks.
When the two stock prices diverge (one stock moves up 
while the other moves down),  the pairs trade would be triggered:
to short the stronger stock and to long the weaker one
betting the eventual convergence of the prices.
The pairs trading was first developed by Bamberger 
and followed by Tartaglia's quantitative group at Morgan Stanley in the 1980s.
A major advantage of pairs trading is its `market neutral' nature
in the sense that it can be profitable under any market conditions.
There are many good discussions in connection with
the cause of the divergence and subsequent convergence. 
We refer the reader to the paper by Gatev et al. \cite{GatevGR},
the book by Vidyamurthy \cite{Vidyamurthy}, and references therein.

In pairs trading, it is important to determine when to initiate
a pairs trade (i.e., how much divergence is sufficient to trigger
a trade) and when to close the position (when to lock in profits
if the stocks perform as expected or when to cut losses if the trade
goes sour). It is the purpose of this paper to 
focus on the mathematics of pairs trading.
In particular, we consider the case when a difference
of a pair satisfies a mean reversion model, follow a dynamic 
programming approach to determine these key thresholds, and 
establish their optimality.

Mean-reversion models are often used in financial 
markets to capture price movements that have the tendency to move
towards an ``equilibrium'' level. There are many studies
in connection with mean reversion stock returns;
see e.g., Cowles and Jones \cite{CowlesJ}) 
Fama and French \cite{FamaF}, and Gallagher and Taylor \cite{GallagherT}
among others. 
In addition to stock markets, mean-reversion
models are also used to characterize stochastic volatility
(Hafner and Herwartz \cite{HafnerH}) and asset prices in energy
markets (see Blanco and Soronow \cite{BlancoS}. 
See also related results in option pricing
with a mean-reversion asset by Bos, Ware and Pavlov \cite{BosWP}.

Mathematical trading rules have been studied for many
years. For example, Zhang \cite{Zhang-trading} considered a
selling rule determined by two threshold levels, a target price
and a stop-loss limit. In \cite{Zhang-trading}, 
such optimal threshold levels are obtained by solving a set of
two-point boundary value problems. 
Guo and Zhang \cite{GuoZ} studied the optimal selling rule under a model with
switching Geometric Brownian motion. 
Using a smooth-fit technique, they 
obtained the optimal threshold levels by solving
a set of algebraic equations.
These papers are concerned with the selling side of trading in
which the underlying price models are of GBM type. 
Recently, Dai et al. \cite{DaiZZ} developed a trend following 
rule based on a conditional probability indicator. 
They showed that the optimal
trading rule can be determined by two threshold curves which
can be obtained by solving the associated Hamilton-Jacobi-Bellman (HJB)
equations.
Similar idea was developed following a confidence interval approach
by Iwarere and Barmish \cite{IwarereB}.
In addition, Merhi and Zervos \cite{MerhiZ} studied 
an investment capacity expansion/reduction problem following
a dynamic programming approach under a geometric Brownian motion market model.
Similar problem under a more general market model was treated by 
L{\o }kka and Zervos \cite{LokkaZ}.
In connection with mean reversion trading, 
Zhang and Zhang \cite{ZhangZ} obtained a buy-low and sell-high
policy by charactering the  `low' and `high' levels in terms of
the mean reversion parameters. 

Despite much progress in various 
mathematical trading rules, an important issue 
hasn't received much attention in the literature: How to cut losses and
how to trade with cutting losses.
In practice, there are many scenarios that cutting losses may arise.
A typical one is margin call.
When the pairs position is undergoing heavy losses,
a margin call may be enforced to close part or the entire position.
In addition, a pairs trader may determine a fixed stop-loss level
from a pure money management consideration.
Furthermore, a historically correlated pairs may cease to be
correlated at some point. For example, acquisition (or bankruptcy) 
of one stock in the pairs position. In this case, it is necessary to
modify the trading rule to accommodate a pre-determined stop-loss level. 
From a control theoretical point of view, adding a stop-loss level 
is amount to impose a hard state constraint. This typically poses 
substantial difficulties in solving the problem.
A major portion of this paper is devoted to address this important issue.

In this paper, we consider an optimal pairs trading rule in which
a pairs (long-short) position consists of
a long position of one stock and a short 
position of the other.
The state process $Z_t$ is defined  as a difference of the 
stock prices.
The objective is to initiate (buy) and close (sell) the pairs positions
sequentially to maximize a discounted payoff function. 
A fixed (commission or slippage) cost will be imposed to each transaction.
As in \cite{ZhangZ}, we
study the problem following a dynamic programming approach and
establish the associated HJB equations
for the value functions. 
We show that the corresponding optimal
stopping times can be determined by three threshold levels $x_0$, $x_1$, 
and $x_2$.
These key levels can be obtained by solving a set of 
algebraic like equations. 
We show that the optimal pairs trading rule can be given in terms
of two intervals: $I_1=[x_0,x_1]$ and $I_2=(M,x_2)$.
Here $M$ is the given stop-loss level (e.g., as the consequence of 
a margin call) and $I_1$ is contained in $I_2$.
The idea to initiate a trade whenever $Z_t$ enters $I_1$ and
hold the position till $Z_t$ exits $I_2$.
In addition, 
we provide a set of sufficient conditions that guarantee the optimality 
of our pairs trading rule.
We also examine the dependence of these threshold levels
on various parameters in a numerical example. Finally, we demonstrate
how to implement the results using a pair of stocks and their
historical prices.

This paper is organized as follows.
In \S2, we formulate the pairs trading problem under consideration.
In \S3, we study properties of the value functions,
the associate HJB equations, and their solutions.
In \S4, we provide a set of sufficient conditions
that guarantee the optimality of our trading rule.
A numerical example is given in \S5. 
The paper is concluded in \S6.

\section{Problem Formulation}
Let $X^1_t$ and $X^2_t$ denote the prices of a pair of correlated stocks
$X^1$ and $X^2$, respectively. 
The corresponding pairs position consists of a long position in stock
$X^1$ and short position in stock $X^2$.
For simplicity, we include one share of $X^1$ and $K_0$ shares of $X^2$
in the pairs position. Here $K_0$ is a given positive number.
The price of the position is given by $Z_t=X^1_t-K_0 X^2_t$.
We assume that $Z_t$ is a mean-reverting (Ornstein-Uhlenbeck) 
process governed by
\beq{sys-X} 
dZ_t=a(b-Z_t)dt+\sigma dW_t,\ Z_0=x, 
\eeq 
where $a>0$ is the rate of reversion, 
$b$ the equilibrium level, $\sigma>0$
the volatility, and $W_t$ a standard Brownian motion. 

In this paper, the notation $X^i$, $i=1,2$, are reserved for the underlying
stocks and $Z$ the corresponding pairs position.
One share long in $Z$ means the combination of 
one share long position in $X^1$ and $K_0$ shares of short position in $X^2$. 
Similarly, for $i=1,2,$, $X^i_t$ represents the price of stock $X^i$ and
$Z_t$ the value of the pairs position at time $t$.
Note that $Z_t$ is allowed to be negative in this paper.

In addition, we impose a state constraint and require
$Z_t\geq M$. Here $M$ is a given constant and it
represents a stop-loss level. 
It is common in practice to limit losses to an acceptable level
to account for unforeseeable events in the marketplace.
A stop-loss limit is often enforced as part of money management.
It can also be associated with a margin call due to substantial
losses. 

To accommodate such state constraint in our model,
let $\tau_M$ denote the exit time of $Z_t$ from $(M,\infty)$, i.e.,
$\tau_M=\inf\{t:\ Z_t\not\in(M,\infty)\}$.


Let
\beq{sequence}
0\leq\tau_1^b\leq\tau_1^s\leq\tau_2^b\leq\tau_2^s\leq\cdots\leq\tau_M
\eeq
denote a sequence of stopping times. A buying decision is made at
$\tau_n^b$ and a selling decision at $\tau_n^s$, $n=1,2,\ldots$.

We consider the case that the net position at any time can be
either long (with one share of $Z$)
or flat (no stock position of either $X^1$ or $X^2$). 
Let $i=0,1$ denote the initial net position. If
initially the net position is long ($i=1$), then one should sell
$Z$ before acquiring any future shares. The corresponding
sequence of stopping times is denoted by
$\Lambda_1=(\tau_1^s,\tau_2^b,\tau_2^s,\tau_3^b,\ldots)$.
Likewise, if
initially the net position is flat ($i=0$), then one should 
start to buy a share of $Z$. The corresponding sequence
of stopping times is denoted by
$\Lambda_0=(\tau_1^b,\tau_1^s,\tau_2^b,\tau_2^s,\ldots)$.

Let $K>0$ denote the fixed transaction cost (e.g., slippage and/or 
commission) associated with buying or selling of $Z$.
Given the initial state $Z_0=x$ and initial net
position $i=0,1$, and the decision sequences,
$\Lambda_0$ and $\Lambda_1$, 
the corresponding reward functions 
\beq{reward-fn}
J_i(x,\Lambda_i)= \left\{\begin{array}{ll} \disp
E\Biggl\{\sum_{n=1}^\infty\left[e^{-\rho\tau_n^s}(Z_{\tau_n^s}-K)
-e^{-\rho\tau_n^b}(Z_{\tau_n^b}+K)\right]\In\Biggr\},&\mbox{ if }i=0,\\
\disp
E\Biggl\{e^{-\rho\tau_1^s}(Z_{\tau_1^s}-K)&\\
\disp\quad
+\sum_{n=2}^\infty\left[e^{-\rho\tau_n^s}(Z_{\tau_n^s}-K)
-e^{-\rho\tau_n^b}(Z_{\tau_n^b}+K)\right]\In\Biggr\},&\mbox{ if }i=1,\\
\end{array}\right.
\eeq
where $\rho>0$ is a given discount factor.

In this paper, given random variables
$\xi_n$, the term $E\sum_{n=1}^\infty \xi_n$ is interpreted as
\[
\limsup_{N\to\infty} E\sum_{n=1}^N \xi_n.
\]

In the reward function $J_i$, a buying decision has to be made before
$Z_t$ reaches $M$. When $t=\tau_M$ (or $Z_t=M$), only a selling 
can be done if $i=1$.

For $i=0,1$, let $V_i(x)$ denote the value functions with the
initial state $Z_0=x$ and initial net positions $i=0,1$. That is,
\beq{value-fn} V_i(x)=\sup_{\Lambda_i}J_i(x,\Lambda_i). \eeq

Note that
\beq{bdry}
V_0(M)=0\mbox{ and }V_1(M)=M-K.
\eeq
These give the boundary conditions.

\begin{rem}\bdd
{\rm 
Note that we allow the equalities in (\ref{sequence}), i.e., 
one can buy and sell simultaneously. Nevertheless, owing to
the existence of positive transactions cost $K$, any simultaneous buying
and selling are automatically ruled out by our optimality conditions.

We also imposed the conditions $\tau^b_n\leq\tau_M$ and $\tau^s_n\leq\tau_M$,
$n=1,2,\ldots$.
If one has a share position of $Z$ and $\tau^s_n=\tau_M$ for some $n$, 
then one has to sell the share to cut losses.
On the other hand,
if $\tau^b_n=\tau_M$, then one should not buy because she has to
sell it right away, which only cause the round trip transaction fees.
}
\end{rem}

\begin{rem}\bdd
{\rm 
Recall that in this paper the stock (pair) price is given by $Z_t$.
In \cite{ZhangZ}, a percentage slippage cost is required and
the stock price is given by $S_t=e^{Z_t}$. Suppose $\widetilde K$ percentage
is added to a buying order. Then the total cost is given by 
$S_t(1+\widetilde K)=e^{Z_t}(1+\widetilde K)$. Its natural logarithm
equals approximately $Z_t+\widetilde K$, which matches the cost 
structure in this paper.
}
\end{rem}

\begin{rem}\bdd
{\rm 
In addition, we only consider the `long' side trading in this paper.
Actually, one can trade by simply reversing the trading rule obtained
in this paper. For example, if $b=0$, then we can trade both
$Z_t$ and $(-Z_t)$ simultaneously because they 
satisfy the same system equation (\ref{sys-X}). 
}
\end{rem}

\begin{rem}\bdd
{\rm 
The optimal stopping problem considered in this paper can be generalized 
to treat similar problems in related fields (e.g., the energy market).
We refer the reader to Hamadene and Zhang \cite{HamadeneZ} and references
therein for additional applications.
}
\end{rem}

\begin{exm}\bdd\label{Example1}
{\rm 
Typically a highly correlated pair can be found from
the same industry sector. In this example, we choose
Wal-Mart Stores Inc. (WMT) and Target Corp. (TGT).
Both companies are from the retail industry and 
they have shared similar dips and highs.
If the price of WMT were to go up a large amount while TGT
stayed the same, a pairs trader would buy TGT and sell short WMT
betting on the convergence of their prices.
In Figure~\ref{WMT-TGT}, the 'normalized' 
(dividing each price by its long term moving average) difference
of WMT and TGT is plotted.
In addition, the data (1992-2012) is divided into two sections. 
The first section 
(1992-2000) is used to calibrate the model and the second section
(2001-2012) to backtest the performance of our results.
Our construction of $Z_t$ determines that the equilibrium level $b=0$. 
By measuring the standard 
derivation of $Z_t$, we obtain the historical volatility 
$\sigma=0.56$. 
Finally, following the traditional least squares method, 
we obtain $a=1.00$.

\begin{center}
\begin{figure}
\begin{center}
\epsfysize=3.5in
\centerline{\epsfbox{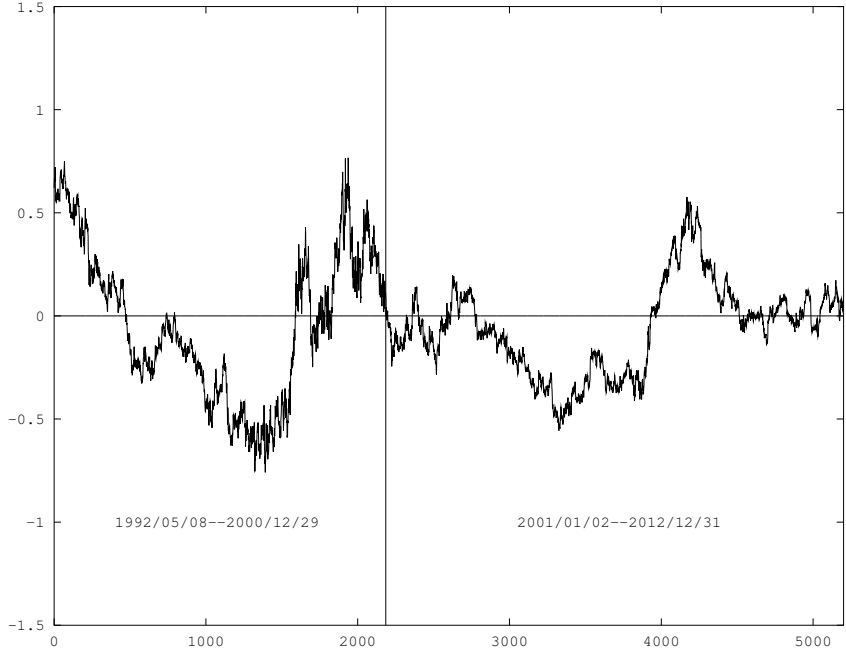}}
\vspace{-3ex}
\nd\caption{{\small WMT and TGT (1992--2012)}}\label{WMT-TGT}
\end{center}
\end{figure}
\end{center}

}
\end{exm}

\section{Properties of the Value Functions}
In this section, we establish various bounds for the value functions
and solve the associated HJB equations.

First, note that the sequence
$\Lambda_0=(\tau^b_1,\tau^s_1,\tau^b_2,\tau^s_2,\ldots)$ can be
regarded as a combination of a buy at $\tau^b_1$ and then
followed by the sequence of stopping times
$\Lambda_1=(\tau^s_1,\tau^b_2,\tau^s_2,\tau^b_3,\ldots)$. In view of
this, we have, for $x>M$, 
\[
\begin{array}{rl}
V_0(x)\geq &\disp J_0(x,\Lambda_0)\\
=&\disp
E\left\{e^{-\rho\tau^s_1}(Z_{\tau^s_1}-K)\Ii+
\sum_{n=2}^\infty\left[e^{-\rho\tau^s_n}(Z_{\tau^s_n}-K)
-e^{-\rho\tau^b_n}(Z_{\tau^b_n}+K)\right]\In\right\}\\
&\disp
-Ee^{-\rho\tau^b_1}(Z_{\tau^b_1}+K)\Ii\\
=&\disp
J_1(x,\Lambda_1)-Ee^{-\rho\tau^b_1}(Z_{\tau^b_1}+K)\Ii.
\end{array}
\] 
In particular, setting $\tau^b_1=0$ and taking supremum over $\Lambda_1$,
we obtain the inequality
\beq{value-ineq-0} V_0(x)\geq V_1(x)-x-K. \eeq
Similarly, we can show, for $x>M$,  that
\beq{value-ineq-1} V_1(x)\geq V_0(x)+x-K. \eeq

Clearly, in view of the boundary conditions (\ref{bdry})
these two inequalities hold for $x=M$.

Next, we establish lower and upper  bounds for $V_i(x)$.

\begin{lem}\label{bounds-value}\bdd
The following inequalities hold:
\[
\begin{array}{l}
0\leq V_0(x)\leq C_0,\\
x-K\leq V_1(x)\leq x+K+C_0,
\end{array}
\]
for all $x\in[M,\infty)$, where $C_0=(\rho+a)|M|/\rho$.
\end{lem}

\nd{\it Proof.}
Note that the lower bounds for $V_i(x)$, ($i=0,1$), 
follow from their definitions.
In addition, if $C_0$ is an upper bound for $V_0(x)$, then 
the upper bound for $V_1(x)$ follows from 
the inequality in (\ref{value-ineq-0}).
It remains to show the upper bound for $V_0$.
Recall that $\tau^b_n\leq\tau^s_n\leq\tau_M$. 
Therefore, we have
\[
E\(W_{\tau^s_n}-W_{\tau^b_n}\)\In
=E\(W_{\tau^s_n}-W_{\tau^b_n}\)-
E\(W_{\tau^s_n}-W_{\tau^b_n}\)I_{\{\tau^b_n=\tau_M\}}
=0.
\]
Recall also that $Z_t\geq M$ for all $t\leq\tau_M$.
Using Dynkin's formula, we have, for each $n=1,2,\ldots$,
\beq{E-e-rho}
\begin{array}{l}
\disp\hspace*{-0.2in}
E\(e^{-\rho \tau_n^s}Z_{\tau_n^s}-e^{-\rho \tau_n^b}Z_{\tau_n^b}\)\In\\
=\disp
E\(\int_{\tau_n^b}^{\tau_n^s}e^{-\rho t}\(-(\rho+a)Z_t\) dt\)\In
+E\(\sigma(W_{\tau^s_n}-W_{\tau^b_n})\)\In\\
\leq\disp
 (\rho+a)|M|E\(\int_{\tau_n^b}^{\tau_n^s}e^{-\rho t} dt\)\In\\
\leq\disp
 (\rho+a)|M|E\int_{\tau_n^b}^{\tau_n^s}e^{-\rho t} dt.
\end{array}
\eeq
It follows from the definition of $J_0(x,\Lambda_0)$ that
\[
\begin{array}{rl}
J_0(x,\Lambda_0)\leq &\disp
\sum_{n=1}^\infty
E\(e^{-\rho \tau_n^s}Z_{\tau_n^s}-e^{-\rho \tau_n^b}Z_{\tau_n^b}\)\Ii\\
\leq &\disp (\rho+a)|M|\sum_{n=1}^\infty 
E\int_{\tau_n^b}^{\tau_n^b}e^{-\rho t}dt\\
\leq &\disp (\rho+a)|M|\int_0^\infty e^{-\rho t}dt\\
=&\disp \frac{(\rho+a)|M|}{\rho}=C_0.
\end{array}
\]
This implies $V_0(x)\leq C_0$.\hfill$\Box$

\vspace*{0.2in}
Let $\A$ denote the generator of $Z_t$, i.e.,
\[
\A=a(b-x)\frac{\partial}{\partial x}+\frac{\sigma^2}{2}
\frac{\partial^2}{\partial x^2}.
\]
Formally, the associated HJB equations should have the form:
\beq{HJB}
\begin{array}{l}
\min\Big\{\rho v_0(x)-\A v_0(x),\ v_0(x)-v_1(x)+x+K\Big\}=0,\\
\min\Big\{\rho v_1(x)-\A v_1(x),\ v_1(x)-v_0(x)-x+K\Big\}=0,\\
\end{array}
\eeq
for $x\in(M,\infty)$, with the boundary conditions
$v_0(M)=0$ and $v_1(M)=M-K$.

\begin{figure}
\setlength{\unitlength}{2.0mm}
\begin{picture}(60,16)(0,0)
\put(0,0){\begin{picture}(60,10)(-30,0)

\put(-20,10){\vector(1,0){60}}

\put(11,9.5){\line(0,1){1}}

\put(11,9.95){\line(1,0){28.5}}
\put(11,10.05){\line(1,0){28.5}}

\put(-20,9.5){\line(0,1){1}}
\put(-20.5,8){{\small $M$}}

\put(-20,9.95){\line(1,0){15.5}}
\put(-20,10.05){\line(1,0){15.5}}

\put(-4.5,9.5){\line(0,1){1}}
\put(-5,8){$x_0$} 

\put(10.5,8){$x_1$} 
\put(-2.5,12){{\small $v_0\!=\!v_1\!-\!x-\!K$}}
\put(20,12){{\small $\rho v_0-\A v_0=0$}}

\put(-18,12){{\small $\rho v_0-\A v_0=0$}}

\put(-20,5){\vector(1,0){60}}

\put(-20,4.95){\line(1,0){38}}
\put(-20,5.05){\line(1,0){38}}

\put(18,4.5){\line(0,1){1}}

\put(-20,4.5){\line(0,1){1}}
\put(-20.5,3){{\small $M$}}

\put(17.5,3){$x_2$} 
\put(-6,2){{\small $\rho v_1-\A v_1=0$}}
\put(23,2){{\small $v_1=v_0\!+\!x\!-\!K$}}

\end{picture}}
\end{picture}
\caption{Continuation Regions}\label{cont-regions}
\end{figure}
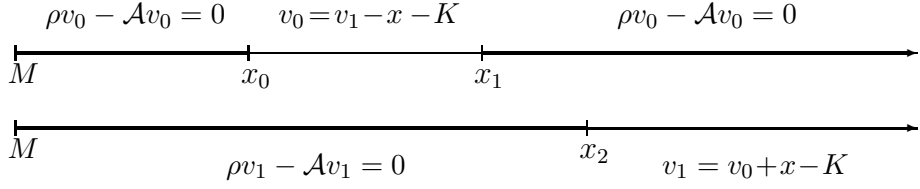

If $i=0$, then one should only buy when the price is low 
(say less than or equal to $x_1$). In this case, $v_0(x)=v_1(x)-x-K$. 
The corresponding continuation region (given by $\rho v_0(x)-\A v_0(x)=0$)
should include $(x_1,\infty)$.
In addition, one should not establish any new position if $Z_t$ is 
close to the stop-loss level $M$. In view of this, the continuation
region should also include $(M,x_0)$ for some $x_0<x_1$.
On the other hand, if $i=1$, then one should only sell 
when the price is high (greater than or equal to $x_2>x_1$), which implies
$v_1(x)=v_0(x)+x-K$ and the continuation region 
(given by $\rho v_1(x)-\A v_1(x)=0$) should be $(M,x_2)$.
These continuation regions are highlighted in Figure~\ref{cont-regions}.

To solve the HJB equations in (\ref{HJB}), we first 
solve the equations $\rho v_i(x)-\A v_i(x)=0$ with
$i=0,1$ on their continuation regions. 
Let 
\[
\left\{\begin{array}{l}
\disp
\phi_1(x)=\int_0^\infty \eta(t)e^{-\kappa(b-x)t} dt,\\
\disp
\phi_2(x)=\int_0^\infty \eta(t)e^{\kappa(b-x)t} dt,\\
\end{array}\right.
\]
where 
$\eta(t)=t^{(\rho/a)-1}\exp\(-t^2/2\)$ and $\kappa=\sqrt{2a}/\sigma$.
Then the general solution (see Eloe et al. \cite{Eloe})
is given by 
$A^0_1\phi_1(x)+A^0_2\phi_2(x)$, for some constants $A^0_1$ and $A^0_2$.

First, consider the interval $(x_1,\infty)$ and 
suppose the solution is given by 
$A_1\phi_1(x)+A_2\phi_2(x)$, for some $A_1$ and $A_2$.
Recall the upper bound for $V_0(x)$ in \lemref{bounds-value},
$v_0(\infty)$ should be bounded above. 
This implies that, $A_1=0$ and $v_0(x)=A_2\phi_2(x)$ on $(x_1,\infty)$.
Let $B_1$, $B_2$, $C_1$, and $C_2$ be constants such that
$v_0(x)=B_1\phi_1(x)+B_2\phi_2(x)$ on $(M,x_0)$ and 
$v_1(x)=C_1\phi_1(x)+C_2\phi_2(x)$ on $(x_2,\infty)$.

It is easy to see that these functions are twice continuously 
differentiable on their continuation regions.
We follow the smooth-fit method which requires the solutions to be
continuously differentiable. 
In particular, it requires $v_0$ to be 
continuously differentiable at $x_0$. Therefore, 
\beq{x0-0}
\left\{\begin{array}{l}
B_1\phi_1(x_0)+B_2\phi_2(x_0)=C_1\phi_1(x_0)+C_2\phi_2(x_0)-x_0-K,\\
B_1\phi_1'(x_0)+B_2\phi_2'(x_0)=C_1\phi_1'(x_0)+C_2\phi_2'(x_0)-1.\\
\end{array}\right.
\eeq

Similarly, the smooth-fit conditions at $x_1$ and $x_2$ yield
\beq{x1-0}
\left\{\begin{array}{l}
A_2\phi_2(x_1)=C_1\phi_1(x_1)+C_2\phi_2(x_1)-x_1-K,\\
A_2\phi_2'(x_1)=C_1\phi_1'(x_1)+C_2\phi_2'(x_1)-1,\\
\end{array}\right.
\eeq
and
\beq{x2-0}
\left\{\begin{array}{l}
C_1\phi_1(x_2)+C_2\phi_2(x_2)=A_2\phi_2(x_2)+x_2-K,\\
C_1\phi_1'(x_2)+C_2\phi_2'(x_2)=A_2\phi_2'(x_2)+1.\\
\end{array}\right.
\eeq

Finally, the boundary conditions at $x=M$ lead to 
\beq{M-0}
\left\{\begin{array}{l}
B_1\phi_1(M)+B_2\phi_2(M)=0,\\
C_1\phi_1(M)+C_2\phi_2(M)=M-K.\\
\end{array}\right.
\eeq

For simplicity in notation, let 
\[
\Phi(x)=\(\begin{array}{cc}
\phi_1(x)&\phi_2(x)\\
\phi_1'(x)&\phi_2'(x)\\
\end{array}\).
\]
Note that the determinant of $\Phi(x)$ is given by 
\[
-\kappa\(
\int_0^\infty\eta(t)e^{-\kappa(b-x)t}dt
\int_0^\infty t\eta(t)e^{\kappa(b-x)t}dt
+\int_0^\infty t\eta(t)e^{-\kappa(b-x)t}dt
\int_0^\infty \eta(t)e^{\kappa(b-x)t}dt\),
\]
which is less than zero for all $x$. Therefore, $\Phi(x)$
is invertible for all $x$.

Also, let
\[
R(x)=\Phi^{-1}(x)
\(\!\!\begin{array}{c}
\phi_2(x)\\
\phi_2'(x)\\
\end{array}\!\!\),\
P_1(x)=\Phi^{-1}(x)
\(\!\!\!\begin{array}{c}
x\!+\!K\\
1\\
\end{array}\!\!\!\),\
P_2(x)=\Phi^{-1}(x)
\(\!\!\!\begin{array}{c}
x\!-\!K\\
1\\
\end{array}\!\!\!\),
\]

Rewrite the equations (\ref{x0-0})-(\ref{M-0}) in terms of 
these vectors. We have
\beq{x0-1}
\(\begin{array}{c}
B_1\\  
B_2\end{array}\)
=
\(\begin{array}{c}
C_1\\  
C_2\end{array}\)
-P_1(x_0),
\eeq

\beq{x1-1}
A_2 R(x_1)=
\(\begin{array}{c}
C_1\\  
C_2\end{array}\)
-P_1(x_1),
\eeq

\beq{x2-1}
\(\begin{array}{c}
C_1\\  
C_2\end{array}\)
=A_2 R(x_2)+P_2(x_2),
\eeq
and
\beq{M-1}
\left\{\begin{array}{l}
(\phi_1(M),\phi_2(M))
\(\begin{array}{c}
B_1\\  
B_2\end{array}\)=0,\\
(\phi_1(M),\phi_2(M))
\(\begin{array}{c}
C_1\\  
C_2\end{array}\)=M-K.
\end{array}\right.
\eeq

Multiplying both sides of (\ref{x0-1})
from the left by $(\phi_1(M),\phi_2(M))$ and using (\ref{M-1}), we have
\beq{x0-2}
(\phi_1(M),\phi_2(M))P_1(x_0)=M-K.
\eeq
Combining (\ref{x1-1}) and (\ref{x2-1}) and eliminating
$(C_1,C_2)^T$, we obtain
\beq{x1x2-0}
A_2(R(x_1)-R(x_2))=P_2(x_2)-P_1(x_1).
\eeq
Also, multiplying both sides of (\ref{x2-1}) from the left by
$(\phi_1(M),\phi_2(M))$ yields
\beq{A2-0}
M-K=A_2(\phi_1(M),\phi_2(M))R(x_2)+(\phi_1(M),\phi_2(M))P_2(x_2).
\eeq
It is easy to check that
\[
(\phi_1(M),\phi_2(M))R(x_2)=\phi_2(M)\det \Phi(x_2)\neq 0.
\]
This leads to 
\beq{A2-1}
A_2=\frac{M-K-(\phi_1(M),\phi_2(M))P_2(x_2)}{(\phi_1(M),\phi_2(M))R(x_2)}.
\eeq
Finally, substitute this into (\ref{x1x2-0}) to obtain
\beq{x1x2}
(R(x_1)-R(x_2))
\(\frac{M-K-(\phi_1(M),\phi_2(M))P_2(x_2)}{(\phi_1(M),\phi_2(M))R(x_2)}\)
=P_2(x_2)-P_1(x_1).
\eeq
Solving equations (\ref{x0-2}) and (\ref{x1x2}), we can obtain
the triple $(x_0,x_1,x_2)$. Then solving the equations
(\ref{x0-1}), (\ref{x1-1}), and (\ref{A2-1}), 
to obtain $A_2$, $(B_1,B_2)$, and $(C_1,C_2)$.


We need additional conditions for $x_1$ and $x_2$. Note that
$v_i(x)$ has to satisfy the following inequalities
for being solutions to the HJB equations (\ref{HJB}):
\beq{additional-conds}
\left\{\begin{array}{l}
\rho v_0(x)-\A v_0(x)\geq0,\\
\rho v_1(x)-\A v_1(x)\geq0,\\
v_0(x)\geq v_1(x)-x-K,\\
v_1(x)\geq v_0(x)+x-K,
\end{array}\right.
\eeq
for all $x\geq M$. Next, we examine each of these inequalities on
intervals $(M,x_0)$, $(x_0,x_1)$, $(x_1,x_2)$, and $(x_2,\infty)$.

First, on $(M,x_0)$, the top two inequalities in (\ref{additional-conds})
become equalities.
We only need the last two inequalities to hold. Therefore, we have
\beq{M-x0}
x-K\leq v_1(x)-v_0(x)\leq x+K\mbox{ on }(M,x_0).
\eeq

Then, on $(x_0,x_1)$, note that $v_0(x)=v_1(x)-x-K$ implies
$v_1(x)\geq v_0(x)+x-K$. We only need $\rho v_0(x)-\A v_0(x)\geq0$.
Again, using  $v_0(x)=v_1(x)-x-K$ and $\rho v_1(x)-\A v_1(x)=0$ on this
interval, we have
\[
\begin{array}{rl}
\rho v_0(x)-\A v_0(x)=&\rho (v_1(x)-x-K)-\A(v_1(x)-x-K)\\
=&\rho (-x-K)-\A(-x-K)\\
=&-(\rho+a)x-\rho K+ab.
\end{array}
\]
In view of this, $\rho v_0(x)-\A v_0(x)\geq0$ on $(x_0,x_1)$ is
equivalent to 
\beq{x1-ineq}
x_1\leq\frac{ab-\rho K}{\rho+a}.
\eeq

Similarly, on $(x_1,x_2)$, we only need the inequalities
\beq{x1-x2-ineq}
x-K\leq v_1(x)-v_0(x)\leq x+K.
\eeq
Finally, on $(x_2,\infty)$, we only require
\beq{x2-infty}
x_2\geq\frac{ab+\rho K}{\rho+a}.
\eeq

Note that the inequalities in (\ref{M-x0}) and (\ref{x1-x2-ineq})
are equivalent to the following inequalities, 
\beq{v0-v1-ineq}
\left\{\begin{array}{ll}
|(C_1-B_1)\phi_1(x)+(C_2-B_2)\phi_2(x)-x|\leq K&\mbox{ on }(M,x_0),\\
|C_1\phi_1(x)+(C_2-A_2)\phi_2(x)-x|\leq K&\mbox{ on }(x_1,x_2),\\
\end{array}\right.
\eeq
respectively.

In what follows, we show that the triple $(x_0,x_1,x_2)$ satisfying these
conditions leads to the optimal stopping rules.

\section{A Verification Theorem}

In this section, we give a verification theorem to show that the
solution $v_i(x)$, $i=0,1$, of equation (\ref{HJB}) are equal to
the value functions $V_i(x)$, $i=0,1$, respectively, and sequences
of optimal stopping times can be constructed from the triple
$(x_0,x_1,x_2)$.

\begin{thm}\label{VerificationThm}\bdd
Let $(x_0,x_1,x_2)$ be a solution to {\rm (\ref{x0-2})} and {\rm (\ref{x1x2})} 
and satisfy
\[
x_1\leq \frac{ab-\rho K}{\rho+a}\mbox{ and }
x_2\geq \frac{ab+\rho K}{\rho+a}.
\]
Let $A_2$, $B_1$, $B_2$, $C_1$, and $C_2$ be constants given by
{\rm (\ref{x0-1})}, {\rm (\ref{x2-1})}, and {\rm (\ref{A2-1})}
satisfying the inequalities in  {\rm (\ref{v0-v1-ineq})}.

Let
\[
\left\{\begin{array}{l}
v_0(x)=\left\{\begin{array}{ll}
\disp
B_1\phi_1(x)+B_2\phi_2(x)&\mbox{ if }x\in[M, x_0),\\
\disp
C_1\phi_1(x)+C_2\phi_2(x)-x-K&\mbox{ if }x\in[x_0, x_1),\\
\disp
A_2\phi_2(x)&\mbox{ if }x\in[x_1, \infty),\\
\end{array}\right.\\
v_1(x)=\left\{\begin{array}{ll}
\disp
C_1\phi_1(x)+C_2\phi_2(x)&\mbox{ if }x\in[M, x_2),\\
\disp
A_2\phi_2(x)+x-K&\mbox{ if }x\in[x_2, \infty).\\
\end{array}\right.\\
\end{array}\right.
\]

Assume $v_0(x)\geq0$. 
Then, $v_i(x)=V_i(x)$, $i=0,1$.
Moreover, if initially $i=0$, let
\[
\Lambda^*_0=(\tau^{b*}_1,\tau^{s*}_1,\tau^{b*}_2,\tau^{s*}_2,\ldots),\\
\]
such that the stopping times 
$\tau_1^{b*}=\inf\{t\geq0:\ Z_t\in[x_0,x_1]\}\wedge\tau_M$,
$\tau^{s*}_n=\inf\{t>\tau^{b*}_n: \ Z_t\not\in(M, x_2)\}\wedge\tau_M$,  and
$\tau^{b*}_{n+1}=\inf\{t>\tau^{s*}_{n}:\ Z_t\in[x_0,x_1]\}\wedge\tau_M$ 
for $n\geq1$.
Similarly, if initially $i=1$, let
\[
\Lambda^*_1=(\sigma^*_1,\tau^*_2,\sigma^*_2,\tau^*_3,\ldots),\\
\]
such that $\tau^{s*}_1=\inf\{t\geq0:\ Z_t\not\in(M,x_2)\}\wedge\tau_M$, 
$\tau^{b*}_{n}=\inf\{t>\tau^{s*}_{n-1}:\ Z_t\in[x_0,x_1]\}\wedge\tau_M$, and
$\tau^{s*}_n=\inf\{t> \tau^{b*}_n: \ Z_t\not\in(M,x_2)\}\wedge\tau_M$ 
for $n\geq 2$. Then $\Lambda^*_0$ and $\Lambda^*_1$ are optimal.
\end{thm}

\nd{\it Proof.} We divide the proof into two steps. 
In the first step, we show that $v_i(x)\geq
J_i(x,\Lambda_i)$ for all $\Lambda_i$. Then in the second step, we
prove that $v_i(x)=J_i(x,\Lambda^*_i)$, which implies $v_i(x)=V_i(x)$
and $\Lambda^*_i$ is optimal.

Let $I_0=(M,x_0)\cup (x_0,x_1)\cup (x_1,\infty)$ and
$I_1=(M,x_2)\cup (x_2,\infty)$. It is easy to see that
$v_0\in C^2(I_0)$, $v_1\in C^2(I_1)$, and
both $v_0$ and $v_1$ are in $C^1([M,\infty))$.
In addition, they satisfy the quasi-variational inequalities in (\ref{HJB}), i.e., $\rho v_i(x)-\A v_i(x)\geq0$, $i=0,1$, whenever they are twice 
continuously differentiable. 
Using these inequalities, Dynkin's formula, and
Fatou's lemma as in ${\O}$ksendal \cite[p. 226]{Oksendal}, we
have, for any stopping times $0\leq \theta_1\leq \theta_2\leq \tau_M$, a.s.,
\beq{two-times-ineq}
\begin{array}{l} 
Ee^{-\rho\theta_1}v_i(Z_{\theta_1})\geq
Ee^{-\rho\theta_2}v_i(X_{\theta_2}),\\
Ee^{-\rho\theta_1}v_i(Z_{\theta_1})I_{\{\theta_1<\tau_M\}}\geq
Ee^{-\rho\theta_2}v_i(X_{\theta_2})I_{\{\theta_1<\tau_M\}}, 
\end{array}
\eeq 
for $i=0,1$. Given
$\Lambda_0=(\tau^b_1,\tau^s_1,\tau^b_2,\tau^s_2,\ldots)$, using
(\ref{value-ineq-0}) and $v_0(M)=0$, we have
\[
\begin{array}{rl}
v_0(x)\geq &\disp
Ee^{-\rho\tau^b_1}v_0(Z_{\tau^b_1})\\
=&\disp
Ee^{-\rho\tau^b_1}v_0(Z_{\tau^b_1})\Ii\\
\geq &\disp
Ee^{-\rho\tau^b_1}\(v_1(Z_{\tau^b_1})-(Z_{\tau^b_1}+K)\)\Ii\\
= &\disp
Ee^{-\rho\tau^b_1}v_1(Z_{\tau^b_1})\Ii-Ee^{-\rho\tau^b_1}(Z_{\tau^b_1}+K)\Ii.
\end{array}
\]
It follows again from (\ref{two-times-ineq}) and then (\ref{value-ineq-1})
that
\[
\begin{array}{rl}
v_0(x) \geq &\disp
Ee^{-\rho\tau^s_1}v_1(Z_{\tau^s_1})\Ii-Ee^{-\rho\tau^b_1}(Z_{\tau^b_1}+K)\Ii\\
\geq &\disp
Ee^{-\rho\tau^s_1}(v_0(Z_{\tau^s_1})+Z_{\tau^s_1}-K)\Ii
-Ee^{-\rho\tau^b_1}(Z_{\tau^b_1}+K)\Ii\\
= &\disp
Ee^{-\rho\tau^s_1}v_0(Z_{\tau^s_1})\Ii+
E\Big[e^{-\rho\tau^s_1}(Z_{\tau^s_1}-K)
-e^{-\rho\tau^b_1}(Z_{\tau^b_1}+K)\Big]\Ii\\
= &\disp
Ee^{-\rho\tau^s_1}v_0(Z_{\tau^s_1})+
E\Big[e^{-\rho\tau^s_1}(Z_{\tau^s_1}-K)
-e^{-\rho\tau^b_1}(Z_{\tau^b_1}+K)\Big]\Ii.\\
\end{array}
\]
Note that
\[
Ee^{-\rho\tau^s_1}v_0(Z_{\tau^s_1})\geq
Ee^{-\rho\tau^b_2}v_0(Z_{\tau^b_2})=Ee^{-\rho\tau^b_2}v_0(Z_{\tau^b_2})
I_{\{\tau^b_2<\tau_M\}}.
\]
Similarly, we have
\beq{tau^b_2}
Ee^{-\rho\tau^s_1}v_0(Z_{\tau^s_1})\geq
Ee^{-\rho\tau^s_2}v_0(Z_{\tau^s_2})+
E\Big[e^{-\rho\tau^s_2}(Z_{\tau^s_2}-K)
-e^{-\rho\tau^b_2}(Z_{\tau^b_2}+K)\Big]I_{\{\tau^b_2<\tau_M\}}.
\eeq

Repeat this process and note that $v_0(x)\geq0$ to obtain
\[
v_0(x)\geq E\sum_{n=1}^N
\Big[e^{-\rho\tau^s_n}(Z_{\tau^s_n}-K)
-e^{-\rho\tau^b_n}(Z_{\tau^b_n}+K)\Big]\In.
\]
Sending $N\to\infty$ to obtain $v_0(x)\geq J_0(x,\Lambda_0)$ for all
$\Lambda_0$. Therefore, $v_0(x)\geq V_0(x)$.

Similarly, using (\ref{tau^b_2}), we can show that 
\[
\begin{array}{rl}
v_1(x)\geq &\disp
Ee^{-\rho\tau^s_1}v_1(Z_{\tau^s_1})\\
\geq &\disp
Ee^{-\rho\tau^s_1}\(v_0(Z_{\tau^s_1})+Z_{\tau^s_1}-K\)\\
= &\disp
Ee^{-\rho\tau^s_1}(Z_{\tau^s_1}-K)+Ee^{-\rho\tau^s_1}v_0(Z_{\tau^s_1})\\
\geq&\cdots\\
= &\disp
Ee^{-\rho\tau^s_1}(Z_{\tau^s_1}-K)
+E\sum_{n=2}^N
\Big[e^{-\rho\tau^s_n}(Z_{\tau^s_n}-K)
-e^{-\rho\tau^b_n}(Z_{\tau^b_n}+K)\Big]\In.
\end{array}
\]
It follows that $v_1(x)\geq V_1(x)$.

Next, we establish the equalities.
Define
$\tau^{b*}_1=\inf\{t\geq0:\ Z_t\in[x_0,x_1]\}\wedge\tau_M$.
Note that $\tau_M<\infty$, a.s. (see \cite[Lemma 6]{ZhangZ}).
Therefore, $\tau^{b*}_1<\infty$, a.s.
Using again Dynkin's formula, we have
\[
\begin{array}{rl}
v_0(x)= &\disp
Ee^{-\rho\tau^{b*}_1}v_0(Z_{\tau^{b*}_1})\\
= &\disp
Ee^{-\rho\tau^{b*}_1}v_0(Z_{\tau^{b*}_1})\Ii\\
= &\disp
Ee^{-\rho\tau^{b*}_1}\(v_1(Z_{\tau^{b*}_1})-(Z_{\tau^{b*}_1}+K)\)\Ii\\
= &\disp
Ee^{-\rho\tau^{b*}_1}v_1(Z_{\tau^{b*}_1})\Ii
-Ee^{-\rho\tau^{b*}_1}(Z_{\tau^{b*}_1}+K)\Ii.\\
\end{array}
\]
Let $\tau^{s*}_1=\inf\{t\geq\tau^{b*}_1:\ X_t=x_2\}\wedge\tau_M$.
Then, $\tau^{s*}_1<\infty$, a.s. We have also
\[
\begin{array}{l}
\disp\hspace*{-0.3in}
Ee^{-\rho\tau^{b*}_1}v_1(Z_{\tau^{b*}_1})\Ii\\
\disp
=Ee^{-\rho\tau^{s*}_1}v_1(Z_{\tau^{s*}_1})\Ii\\
\disp
=Ee^{-\rho\tau^{s*}_1}\(v_0(Z_{\tau^{s*}_1})+(Z_{\tau^{s*}_1}-K)\)\Ii\\
\disp
=Ee^{-\rho\tau^{s*}_1}v_0(Z_{\tau^{s*}_1})\Ii+Ee^{-\rho\tau^{s*}_1}
(Z_{\tau^{s*}_1}-K)\Ii\\
\disp
=Ee^{-\rho\tau^{s*}_1}v_0(Z_{\tau^{s*}_1})+Ee^{-\rho\tau^{s*}_1}
(Z_{\tau^{s*}_1}-K)\Ii.\\
\end{array}
\]
It follows that
\[
v_0(x)=Ee^{-\rho\tau^{s*}_1}v_0(Z_{\tau^{s*}_1})+
E\Big[e^{-\rho\tau^{s*}_1}(Z_{\tau^{s*}_1}-K)
-e^{-\rho\tau^{b*}_1}(Z_{\tau^{b*}_1}+K)\Big]\Ii.
\]
Continue this way to obtain
\[
v_0(x)=Ee^{-\rho\tau^{s*}_N}v_0(Z_{\tau^{s*}_N})+
E\sum_{n=1}^N \Big[e^{-\rho\tau^{s*}_n}(Z_{\tau^{s*}_n}-K)
-e^{-\rho\tau^{b*}_n}(Z_{\tau^{b*}_n}+K)\Big]\IIn.
\]

Similarly,  we can show
\[
\begin{array}{rl}
v_1(x)=&\disp
Ee^{-\rho\tau^{s*}_1}v_1(Z_{\tau^{s*}_1})\\
=&\disp
Ee^{-\rho\tau^{s*}_1}(v_0(Z_{\tau^{s*}_1})+Z_{\tau^{s*}_1}-K)\\
=&\disp
Ee^{-\rho\tau^{s*}_1}v_0(Z_{\tau^{s*}_1})+
Ee^{-\rho\tau^{s*}_1}(Z_{\tau^{s*}_1}-K)\\
=&\disp
Ee^{-\rho\tau^{s*}_N}v_0(Z_{\tau^{s*}_N})+
Ee^{-\rho\tau^{s*}_1}(Z_{\tau^{s*}_1}-K)\\
&\disp
+E\sum_{n=2}^N \Big[e^{-\rho\tau^{s*}_n}(Z_{\tau^{s*}_n}-K)
-e^{-\rho\tau^{b*}_n}(Z_{\tau^{b*}_n}+K)\Big]\IIn.
\end{array}
\]

Recall that $P(\tau_M<\infty)=1$. This implies 
$\lim_{N\to\infty}\tau^{s*}_N=\tau_M$, a.s.
Recall also that $v_0(M)=0$. It follows that
$Ee^{-\rho\tau^{s*}_n}v_0(Z_{\tau^{s*}_n})\to0$.
This completes the proof. $\Box$

\section{A Numerical Example}

In this section, we use the parameters of the WMT-TGT example, i.e.,
\[
a=1.0,\
b=0,\
\sigma=0.56,\
\rho=0.10,\
K=0.001.
\]
Solving the equations (\ref{x0-2}) and (\ref{x1x2}) gives
the triple $(x_0,x_1,x_2)=(-0.142,-0.077,0.077)$.
Next, we vary one of the parameters at a time and examine 
the dependence of the triple $(x_0,x_1,x_2)$ on these parameters.

\subsection*{Dependence of $(x_0,x_1,x_2)$ on parameters}
First we consider the triple $(x_0,x_1,x_2)$ associated with varying $a$. 
A larger $a$ implies larger pulling rate back to the equilibrium $b=0$.
It can be seen in Table 1 that the lower buying level $x_0$ decreases
as $a$ gets bigger.
Also the higher buying level $x_1$ increases in $a$.
These lead to larger buying interval $[x_0,x_1]$ resulting
greater buying opportunities.
The selling level $x_2$ decreases which
suggests one should take profit sooner as $a$ gets bigger because the
potential of going higher becomes smaller. In addition, the interval
$(x_1,x_2)$ is symmetric about $b=0$. 

{\small
\begin{center}
\begin{tabular}{|c|c|c|c|c|c|}\hline
$a$&0.60&0.80&1.00&1.20&1.40\\ \hline
$x_0$&-0.124&-0.135&-0.142&-0.147&-0.151\\ \hline
$x_1$&-0.089&-0.083&-0.077&-0.073&-0.069\\ \hline
$x_2$& 0.089& 0.083& 0.077& 0.073& 0.069\\ \hline
\end{tabular}

\vspace*{0.1in}
\centerline{Table 1. $(x_0,x_1,x_2)$ with varying $a$.}
\end{center}
}

In Table 2, we vary the volatility $\sigma$.
The volatility is the source forcing the price to go away from
its equilibrium. The large the $\sigma$, the further the price 
fluctuates. As a result, every element in the triple $(x_0,x_1,x_2)$
moves along the opposite direction as $\sigma$ increases resulting
a smaller buying interval $[x_0,x_1]$ and a higher profit target $x_2$.

{\small
\begin{center}
\begin{tabular}{|c|c|c|c|c|c|}\hline
$\sigma$&0.36&0.46&0.56&0.66&0.76\\ \hline
$x_0$&-0.164&-0.153&-0.142&-0.130&-0.117\\ \hline
$x_1$&-0.057&-0.067&-0.077&-0.086&-0.095\\ \hline
$x_2$& 0.057& 0.067& 0.077& 0.086& 0.095\\ \hline
\end{tabular}

\vspace*{0.1in}
\centerline{Table 2. $(x_0,x_1,x_2)$ with varying $\sigma$.}
\end{center}
}

Next, we vary the discount rate $\rho$. Larger $\rho$ means 
quicker profits. This is confirmed in Table 3. It shows 
that larger $\rho$ leads to a smaller $x_0$, a slightly larger $x_1$,
and a slightly smaller $x_2$.
This means more buying opportunities and quicker profit taking.

{\small
\begin{center}
\begin{tabular}{|c|c|c|c|c|c|}\hline
$\rho$&0.06&0.08&0.10&0.12&0.14\\ \hline
$x_0$&-0.1412&-0.1416&-0.1420&-0.1426&-0.1430\\ \hline
$x_1$&-0.078&-0.078&-0.077&-0.077&-0.076\\ \hline
$x_2$& 0.078& 0.078& 0.077& 0.077& 0.076\\ \hline
\end{tabular}

\vspace*{0.1in}
\centerline{Table 3. $(x_0,x_1,x_2)$ with varying $\rho$.}
\end{center}
}

Finally, we examine the dependence on the stop-loss level $M$. 
Clearly, a smaller $M$ is associated with a larger loss
when it goes wrong. In Table 4, the lower buying level $x_0$ decreases
in $M$. On the other hand, the buying-selling interval 
$(x_1,x_2)$ is not as sensitive to variations in $M$.

{\small
\begin{center}
\begin{tabular}{|c|c|c|c|c|c|}\hline
$M$&-0.16&-0.18&-0.20&-0.22&-0.24\\ \hline
$x_0$&-0.091&-0.118&-0.142&-0.166&-0.189\\ \hline
$x_1$&-0.077&-0.078&-0.077&-0.077&-0.077\\ \hline
$x_2$& 0.077& 0.078& 0.077& 0.077& 0.077\\ \hline
\end{tabular}

\vspace*{0.1in}
\centerline{Table 4. $(x_0,x_1,x_2)$ with varying $M$.}
\end{center}
}

\subsection*{Backtesting (WMT-TGT)}
\begin{center}
\begin{figure}
\begin{center}
\epsfysize=3.5in
\centerline{\epsfbox{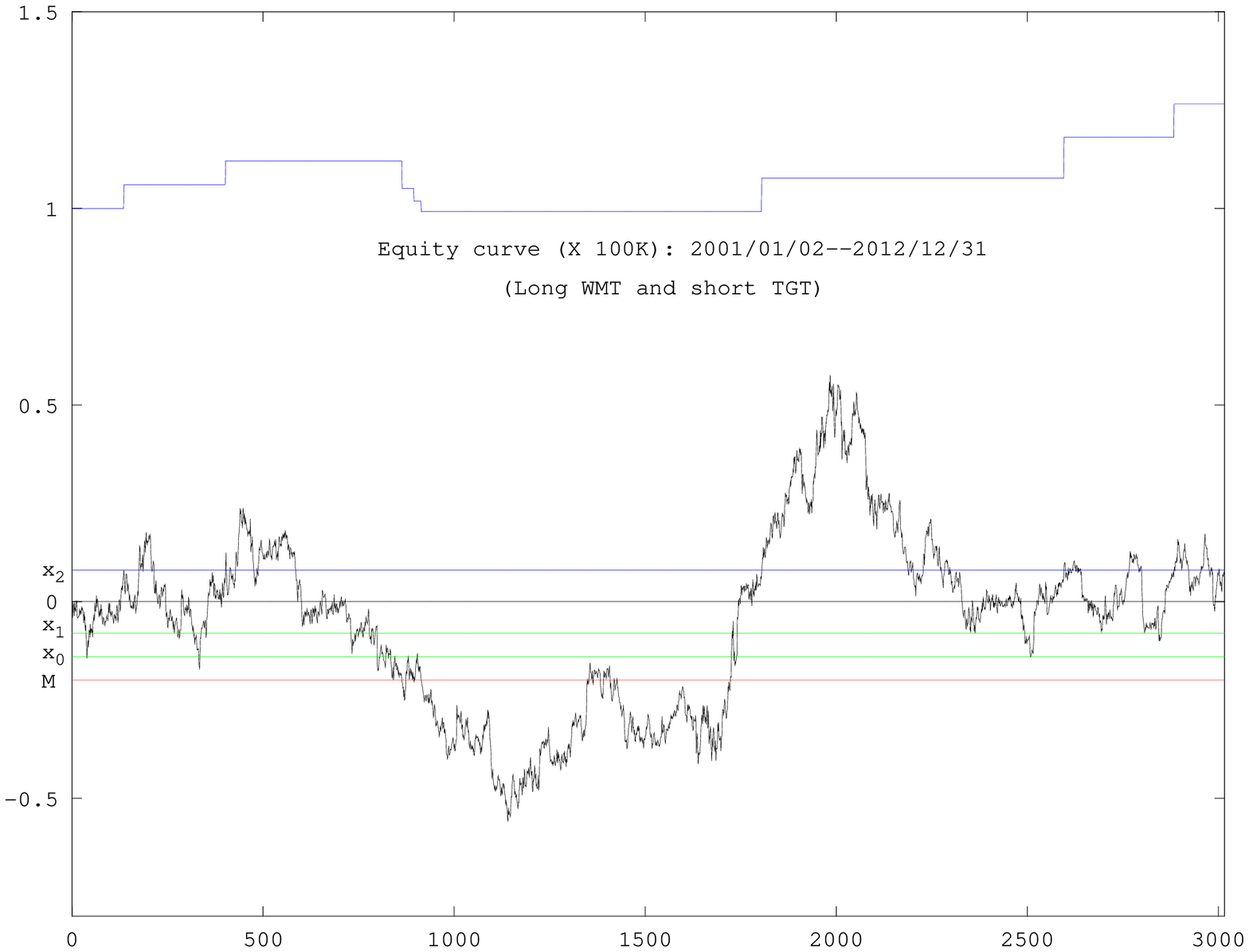}}
\vspace{-3ex}
\nd\caption{{\small Threshold levels and the equity curve}}\label{eq-WMT-TGT}
\end{center}
\end{figure}
\end{center}

We backtest the pairs trading rule using the stock prices
of WMT and TGT from 2001 to 2012. 
Let $X^1_t$ be the WMT stock divide by its 1000 day moving average and 
$X^2_t$ the TGT stock by its same period moving average.
We take $Z_t=X^1_t-X^2_t$. Using the parameters obtained 
in \exmref{Example1}
based on the historical prices from 1992 to 2000, we found
the triple $(x_0,x_1,x_2)=(-0.142,-0.077,0.077)$. 
A pairs trading is triggered when $Z_t$ gets inside the buying
interval $[x_0,x_1]$. The position is closed when $Z_t$ exits
the interval $(M,x_2)$.
Initially, we allocate trading the capital $\$$100K. 
When the first long signal is triggered, buy $\$$50K WMT stocks and
short the same amount TGT. Close the position either when $Z_t$
reaches the target $x_2$ or when it drops below the stop-loss level $M$.
Such half-and-half capital allocation between long and short applies
to all trades.
In addition, each pairs transaction is charged $\$$5 commission fee. 
Furthermore, two variations from the assumptions prescribed in 
\thmref{VerificationThm} in our `actual' trading:
(a) After the stop-loss level $M$ is reached, the trading 
continues and a buying order is entered 
when $Z_t$ goes back to the trading range;
(b) All available capital will be used (half long and half short) for trading
rather than following the `single' share rule, 

In Figure~\ref{eq-WMT-TGT}, the corresponding $Z_t$, the threshold triple,
and the corresponding equity curve are plotted. There are total 8 trades
and the end balance is $\$$126.602K.

\begin{center}
\begin{figure}
\begin{center}
\epsfysize=3.5in
\centerline{\epsfbox{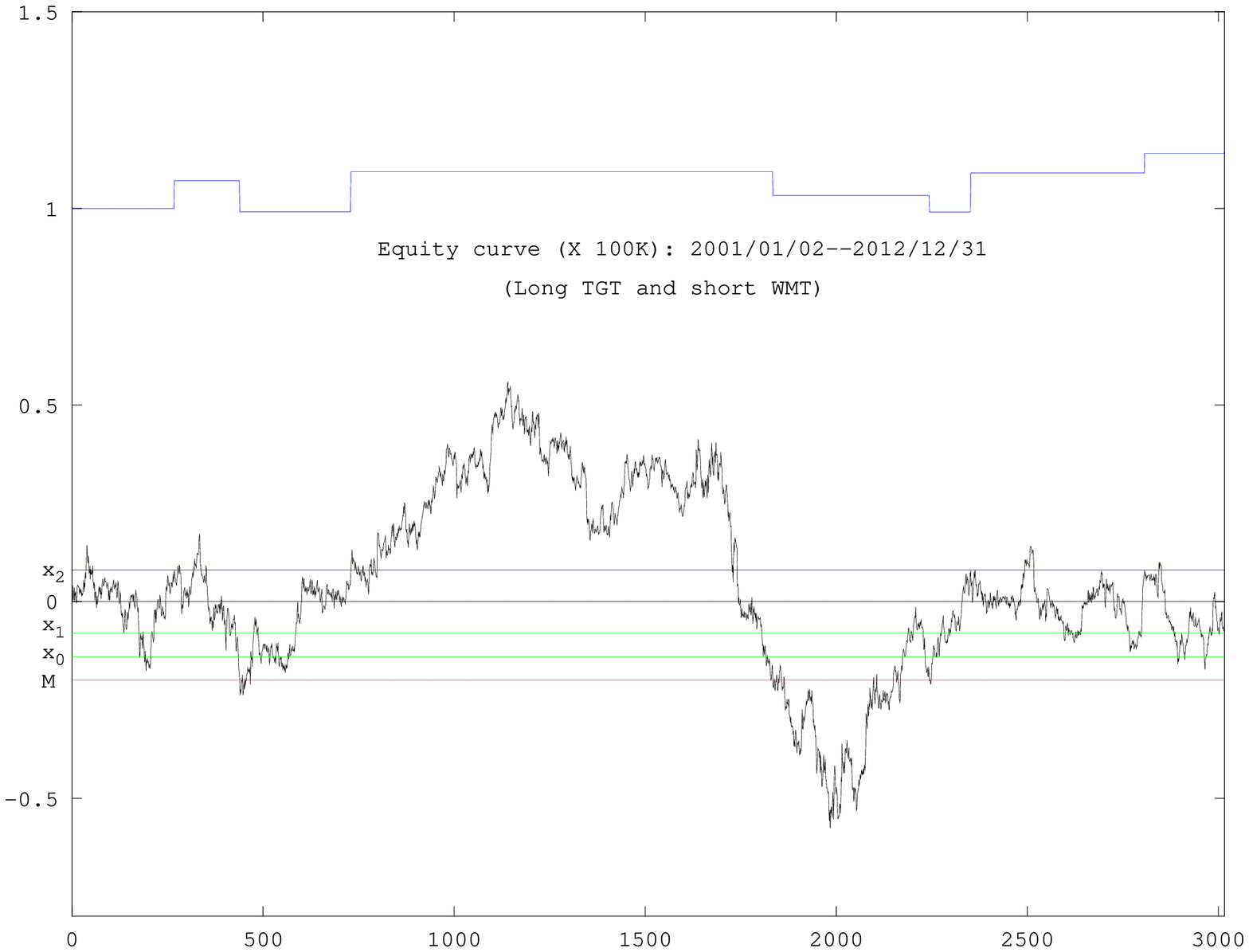}}
\vspace{-3ex}
\nd\caption{{\small Threshold levels and the equity curve}}\label{eq-TGT-WMT}
\end{center}
\end{figure}
\end{center}

Note that $Z_t$ is symmetric, i.e., $(-Z_t)$ satisfies the same
equation (\ref{sys-X}). Naturally, one can reverse the pair and
trade $(-Z_t)$ the same way. The reversed $Z_t$ and equity curve is 
given in Figure~\ref{eq-TGT-WMT}.
Such trade leads to the end balance $\$$114.935K. 
Note that both types of trades have no overlap, i.e., they do not 
compete for the same capital. The grand total profit is
$\$$41547 which is a \%41.54 gain. 

The main advantage of pairs trading is its risk neutral nature, i.e.,
it can be profitable regardless the general market condition.
In addition, there are only 2x8 trades leaving the capital
in cash most of the time. This is desirable because the cash 
sitting in the account can be used for other types of shorter term
trading in between, at least drawing interest over time.

Finally, the choice of stop-loss level $M$ can depend on many factors
including the trader's risk tolerance level and margin requirements.
Our choice $M=-0.2$ corresponds to a \%10 loss when 
WMT drops \%10 and TGT stays the same. 

\section{Conclusion}
In this paper, we have studied the pairs trading problem
following a mean reversion approach and 
obtained a closed-form solution under reasonable conditions.
Much attention was given to the trading rule with loss cutting, which is an
important component of money management. 

A simple real market (WMT-TGT) example was considered.
It would be interesting to examine how the method works for a 
larger selection of pairs of correlated stocks. Some practical
considerations can be found in the book by Vidyamurthy \cite{Vidyamurthy}.


\end{document}